\documentclass[a4paper,usenatbib]{mnras}

\usepackage{newtxtext,newtxmath}

\usepackage[T1]{fontenc}
\usepackage{ae,aecompl}

\makeatletter

\newcommand{\Rmnum}[1]{\expandafter\@slowromancap\romannumeral #1@}
\makeatother

\usepackage{graphicx}	
\usepackage{amsmath}	
\usepackage{amssymb}	
\usepackage{subcaption}
\usepackage{color}

\title[Quasar gas outflows at z=7.54 from BlueTides]{Gas outflows from the z= 7.54 quasar: predictions from the BlueTides Simulation.}

\author[et al.]{Yueying Ni$^{1}$\thanks{\tt yueyingn@andrew.cmu.edu}, Tiziana Di Matteo$^{1}$\thanks{\tt tiziana@phys.cmu.edu}, Yu Feng$^{2}$, Rupert A.C. Croft$^{1}$
and Ananth Tenneti$^{1}$\\
$^{1}$ McWilliams Center for Cosmology, Department of Physics, Carnegie Mellon University, Pittsburgh, PA 15213, USA\\
$^{2}$ Berkeley Center for Cosmological Physics, University of California, Berkeley, CA 94720, USA
}

\date{Accepted XXX. Received YYY; in original form ZZZ}

\pubyear{2018}

\begin{document}
\label{firstpage}
\pagerange{\pageref{firstpage}--\pageref{lastpage}}
\maketitle

\begin{abstract}
Many theoretical models predict that quasar driven outflows account  for the observed quenching of star formation in massive galaxies. There is growing observational evidence for quasar-launched massive outflows at z>6, while the details of outflow-host galaxy interaction remain obscure. In this paper, we study the feedback around the highest redshift quasar in the BlueTides simulation, the largest volume cosmological hydrodynamic simulation so far carried out. We present predictions for gas outflows around the
brightest $z = 7.54$ 
quasar  which hosts the most massive black hole in the simulation volume, which has grown to black hole mass $6.7\times 10^{8}{\rm M}_\odot$ consistent with the current record holder for high-z quasars.
We introduce a method to identify and trace the gas outflowing from the halo. We find that the total mass of the outflow gas is about $3.6\times 10^{9}{\rm M}_\odot$, constituting 6\% of the total gas in the halo. 
The outflow gas contains a cold, dense molecular component with mass about $2.6\times 10^{8}{\rm M}_\odot$, that originates from the inner region of the halo, within  a few kpc of the central black hole. The velocities of the outflow gas reach thousands of km/s, within which the molecular component has mass averaged outward radial velocity of $1300$ km/s, consistent with observations. The averaged outflow rate is about $200-300 {\rm M}_\odot/yr$, with the outflowing gas mainly in a hotter ($T \sim 10^7$ K) and lower density state than the average of the host halo gas. 
\end{abstract}

\begin{keywords}
methods: numerical - galaxies: evolution - quasars: supermassive black holes
\end{keywords}


\section{Introduction}
High redshift quasars are powerful tracers of black hole growth and galaxy formation in the early Universe. 
Surrounded by gas of high density, the accretion rate of a quasar can reach near Eddington levels, leading to extremely high luminosity. 
Observed scaling relation between SMBH mass and properties (e.g, the bulge mass, the central velocity dispersion) of the host galaxy indicate a strong connection between the supermassive black hole in the center of the galaxy and the evolution of its host~\citep{Silk1998,Kormendy}. 
The feedback from AGN appears to be essential for galaxy evolution, being a good candidate to quench star formation in early galaxies~\citep{Fabian2012,Cheung}. 
The high radiation output of quasars can couple to the surrounding gas, heating up and launching powerful hot gas outflows. 
A giant outflow may halt the global inflow of cooling halo gas from filamentary accretion and thus restrict the fuel for star formation. 
Outflow gas may also couple to cold dense molecular gas in the galactic disk and quench the star formation by directly expelling a fraction of the star forming gas from the host galaxy. 

Discovery of old, passive and massive galaxies at $z\sim2$ ~\citep[e.g.][]{Cimatti2004, Saracco2005} implies that the quasar driven feedback quenching mechanism must have been at work at high-z. 
Observational evidence for massive quasar-driven outflows at high redshifts(z>6) has already been found. 
In the host galaxy of SDSS J1148+5251 at z=6.4, a powerful gas outflow has been detected ~\citep{Maiolino} through the presence of extremely broad wings of the [C \Rmnum{2}] emission line extending from $-1400$ km/s to $1200$ km/s from the systemic velocity, resolved on scale of $r \sim 16$ kpc.
\cite{Cicone2015} further indicated that the outflow has a complex morphology in cold dense molecular gas, which is distributed up to $r \sim 30$ kpc from the center of the host galaxy, and revised the lower limit of the mass outflow rate to $1400 \rm{M}_{\odot}$/yr. 

The details of the interaction between outflow and host galaxy, and how the cold, molecular component couples to the hot outflow gas still remains unclear. 
Recent simulation studies ~\citep[e.g.][]{Costa, Curtis, Biernacki, Barai2018} show that dense and cold gas can emerge from galaxies at high velocity, but so far do not include direct prescriptions for the molecular
gas component in the simulation.
The claim is made that the powerful outflows of dense, cold gas only occur when supernova and AGN driven outflows work in tandem. \cite{Costa} deduced that the cold gas comes from radiative cooling of the hot AGN-driven outflow when propagating through a clumpy medium pre-enriched with metals from SN. ~\cite{Biernacki} inferred that the cold gas in outflow originates from the metal-enriched SN-driven galactic fountain, and is then further accelerated by the hot AGN-driven outflow so that it sweeps through the full extent of halo.

Observations of quasar J1342+0928 with black hole mass $8\times 10^{8}\rm{M}_\odot$ at z=7.54 have been reported in \cite{Banados}. This is the highest redshift quasar yet observed, residing in a crucial stage of reionization epoch. ~\cite{Venemans2017} further reported the detection of copious amounts of dust and [C II] emission from the interstellar medium (ISM) of its host galaxy.

The outflow of a quasar at this high redshift ($z>7$) has not yet been detected. In this paper we present prediction on the gas outflow in the host halo of the most massive quasar in BlueTides simulation, covering the epoch from z=8 to z=7.54. The mass of the central blackhole has grown to $6.7\times 10^{8}\rm{M}_\odot$ by z=7.54, quite close to the value for the newly observed earliest quasar. We study the gas outflow with large radial velocity in the simulation at this (currently unexplored) high redshift and directly track the cold molecular phase in the outflow.

The paper is organized as follows: we introduce the Bluetides simulation code and our method to determine which are the outflowing gas particles in section 2, present our results for outflow properties in section 3, and give summary of our findings in section 4.

\begin{figure}
\includegraphics[width=\columnwidth]{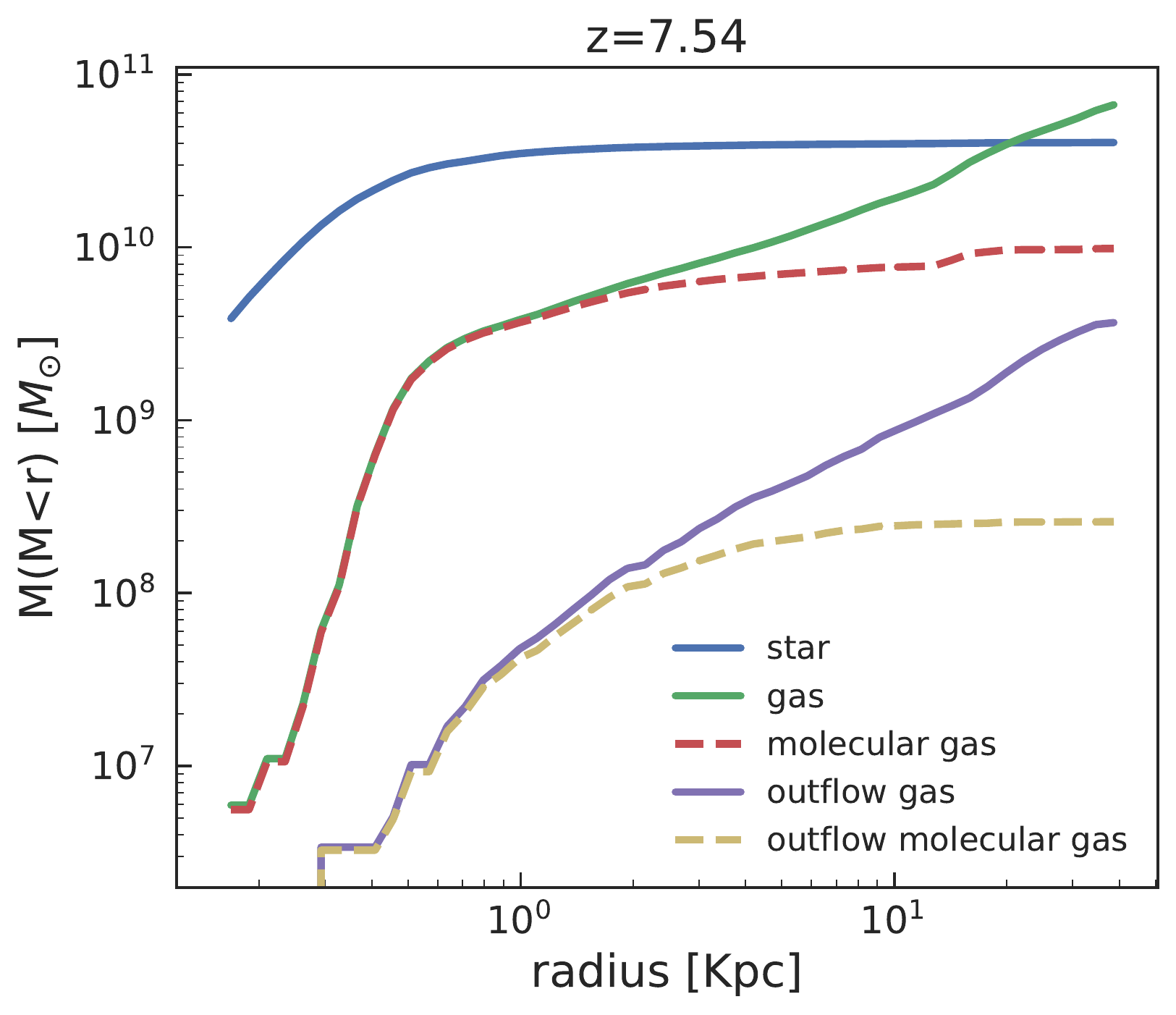}
\caption{The cumulative mass profile of the star, gas and the molecular gas component in the halo hosting the $z=7.54$ quasars. Red and yellow dashed lines represent the molecular component in the gas profile and outflow gas profile respectively (discussed in \S2). }
\label{fig:figure01}
\end{figure}

\section{Methods}

\subsection{BlueTides simulation}

The BlueTides cosmological simulation~\citep{feng2016} uses the Pressure Entropy Smoothed Particle Hydrodynamics code MP-Gadget to follow the evolution of a $(400h^{-1}\rm{Mpc})^{3}$ cubical volume containing $2\cdot7040^{3}$ particles. The initial conditions are at $z=99$ and the run has reached below $z=7.5$. The large volume and high resolution of BlueTides, almost 200 times larger than either the Illustris~\citep{Vogelsberger}, or EAGLE~\citep{Schaye}, simulations, makes it ideally suited to study rare objects at high-redshift, along with the detailed structure and physical properties of their internal and surrounding gas.

BlueTides implements a variety of sub-grid models for physical processes so that their effects on galaxy formation can be studied. 
Star formation is based on a multi-phase star formation model ~\citep{Springel01} with modifications following~\cite{2013MNRAS.436.3031V}.
Gas is allowed to cool through both radiative processes~\citep{Katz} and metal cooling~\citep{Vogelsberger}. 
Type II supernova wind feedback (the model used in Illustris ~\citep{Nelson}) is included, assuming wind speeds proportional to the local one dimensional dark matter velocity dispersion. 
The large volume of BlueTides also allows to include a model of "patchy reionization" ~\citep{Battaglia}, yielding a mean reionization redshift $z\approx10$, and incorporating the UV background estimated by \cite{Faucher}. 

The feedback from AGN is modeled in the same way as for the MassiveBlack I and II simulations, which results in self-regulated super-massive black hole accretion following~\citep{DiMatteo}. 
Super-massive black holes are seeded with an initial mass of $5 \times 10^5 h^{-1}\rm{M}_{\odot}$ in halos more massive than $5 \times 10^{10} h^{-1}\rm{M}_{\odot}$. 
The bolometric luminosity of AGN is $L = \eta \dot{M}c^2$, with $\eta = 0.1$ being the mass-to-light conversion efficiency in the accretion process.
5\% of the radiation energy is thermally coupled to the surrounding gas as feedback energy, deposited in a sphere of twice the radius of the SPH smoothing kernel of the black hole. 
The AGN feedback energy only appears in the kinetic form through this thermal energy deposition, and no other coupling (e.g.,  radiation presssure) is included.

We model the formation of molecular hydrogen and its effect on star formation at low metallicities following the prescription of \cite{Krumholtz}. 
Since our work directly predicts the molecular gas component in the outflow, we briefly review the sub-grid physics used in the simulation for estimating  the $\rm{H}_2$ mass fraction here.   
The approach starts with a spherical gas cloud immersed in a uniform Lyman-Werner band radiation field and assumes the cloud is in steady state to solve the radiative transfer and $\rm{H}_2$ formation-dissociation balance equations. 
After some approximations, \cite{McKee} found the result
\begin{equation}
f_{\rm{H}_2} = 1 - \frac{3}{4}\frac{s}{1+0.25s}
\end{equation}
where $f_{\rm{H}_2}$ is the mass fraction of molecular hydrogen. The parameter $s$ represents the size of the atomic-molecular complex that can be calculated to be
\begin{equation}
s=\frac{\ln(1+0.6\chi+0.01\chi)}{0.6 \tau_{c}}
\end{equation}
where $\tau_{c}$ is the dust optical depth of the atomic-molecular cloud and $\chi$ can be thought as an estimate of the local radiation field. 
These two parameters can be calculated to be
\begin{equation}
\tau_{c}= \frac{\Sigma_{\rm{H}}\sigma_{d}}{\mu_{\rm{H}}} \;;\;\;\; 
\chi \approx 3.1\frac{1+3.1 (Z/Z_{\odot})^{0.365}}{4.1}
\end{equation}

Here, $\Sigma_{\rm{H}} \approx \rho_{\rm{H}}^2/|\nabla\rho_{\rm{H}}|$ is the local hydrogen column density determined using a Sobolev-like approximation, $\sigma_{d}$ is the dust cross section per hydrogen nucleus and $\mu_{\rm{H}}=2.3 \times 10^{-24}$g is the mean mass per H nucleus. 
The dust cross section is taken to be $\sigma_{d,-21} = 10^{-21} (Z/Z_{\odot})(\rm{cm}^2)$, and $Z_{\odot}=0.02$ is the gas metallicity of the solar neighborhood. 
With these equations we are able to calculate the $\rm{H}_2$ fraction of each gas particle.
 
In the BlueTides simulation, we identify the halos using the Friends-of-Friends algorithm with a linking length of 0.2 times the mean particle separation.

\begin{figure*}
\includegraphics[width=2.3\columnwidth]{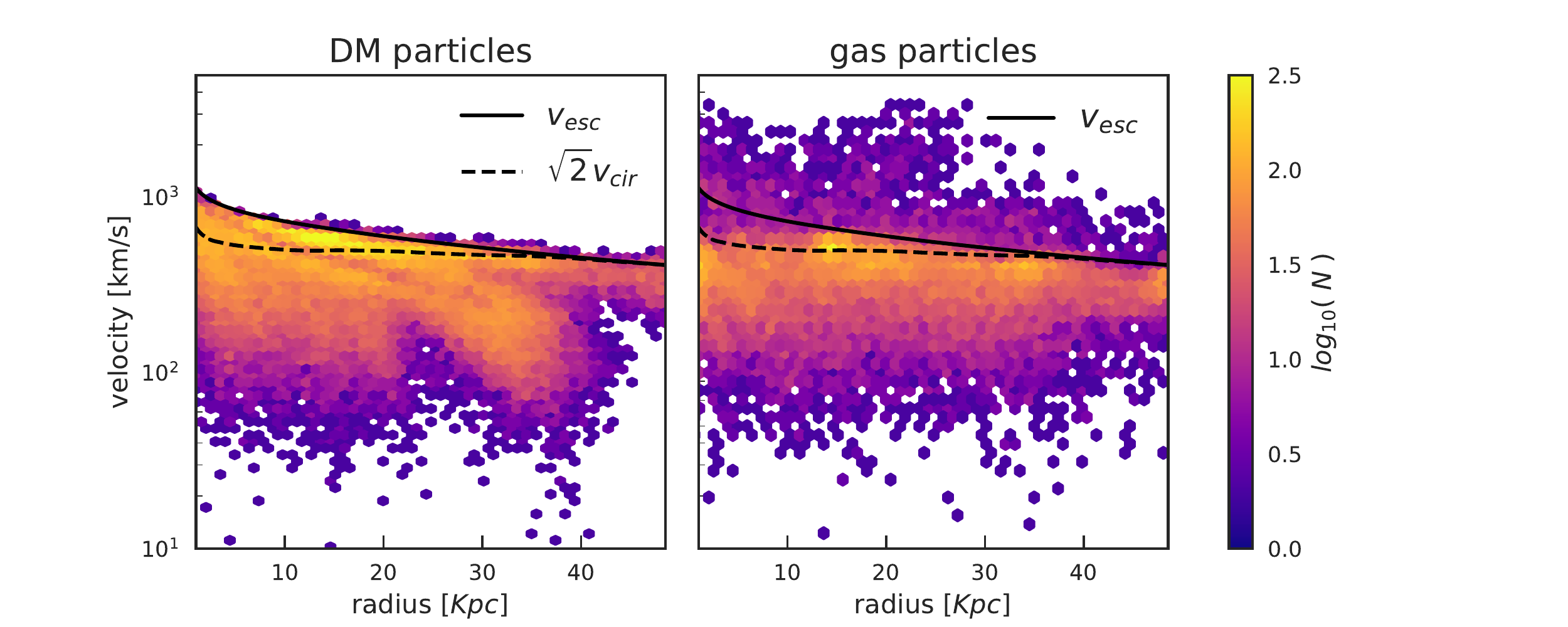}
    \caption{The velocity distribution of dark matter (left panel) and gas particles (right panel) at $z=7.54$. The escape velocity profile (black line) aligns with dark matter particles well. The gas particles whose velocity exceed the escape velocity are outflowing. We also plotted using a dashed line $v(r)=\sqrt{2}v_{\rm cir}=\sqrt{2GM(<r)/r}$ to show the difference between this and our calculated $v_{\rm esc}$.}
    \label{fig:figure2}
\end{figure*}

With the large volume and high resolution of BlueTides, we are able to model the formation of bright quasars that arise in extremely rare high-density peaks in the very early universe.
The most massive black hole formed in the BlueTides simulation has gone through extremely rapid growth at early epochs and has reached a mass that is within 20\% of that powering the recently discovered quasar with the highest redshift ($z=7.54$) \citep{Banados}.
Work on an earlier stage ($z>8$) of the BlueTides simulation has been done to understand what leads to this vigorous growth of the massive quasar in early universe and probe the environment of the host galaxy~\citep{DiMatteo2017}. 
In this paper we focus on the feedback effect of this quasar and we investigate whether outflows are already self-regulating the growth of these extreme black holes/quasars at these early epochs.
In particular we will study the gas outflows from redshifts $z=8$ to $z=7.54$. 

Figure \ref{fig:figure01} shows the cumulative mass profile of stars and gas, and also the molecular component of the gas of the host galaxies for the quasar.

At z=7.54, the mass of the halo is $M_{200}=6.3\times 10^{11}{\rm M}_\odot$, with the virial radius being $R_{200}=36$ kpc. The mass of the central blackhole is $6.7\times 10^{8}{\rm M}_\odot$.  The total mass of the gas in the halo is $6.4\times10^{10}{\rm M}_\odot$ (this is the total gas mass out to $R_{200}$). 
More detailed predictions for the host galaxy of this quasar are given in an associated paper ~\citep[][see also Table 1]{Tenneti2018}.

\begin{figure}
\includegraphics[width=\columnwidth]{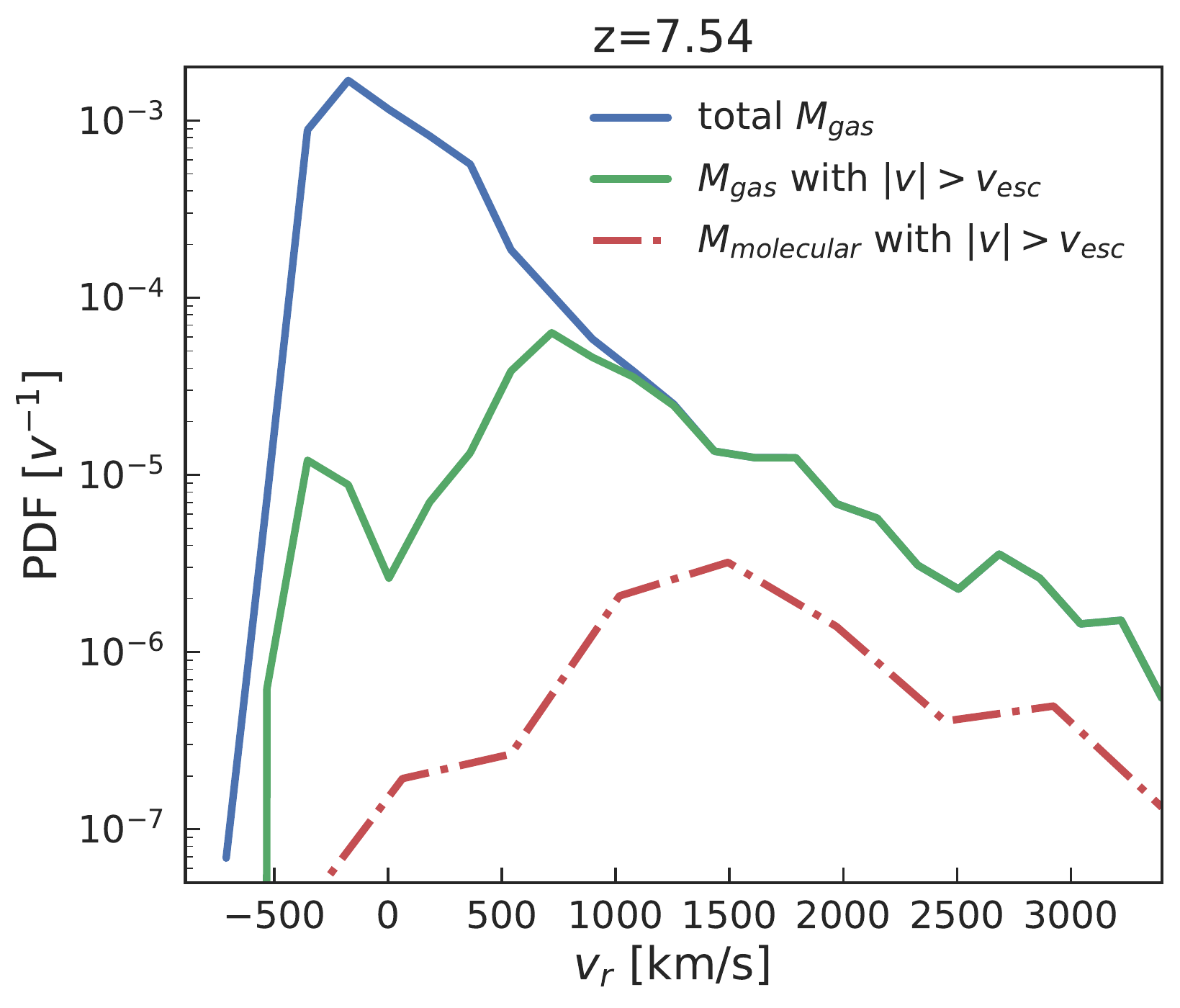}
\caption{The mass weighted histogram of radial velocity of the gas particles at z=7.54. The blue line shows all the gas in the halo. The integral of the blue line is normalized to 1. The green line represents the particles with speed larger than escape velocity (outflowing) and the red dashed line corresponds to the molecular component of the green line.}
\label{fig:figure3}
\end{figure}

\subsection{Gas Outflows}
We introduce a way to identify the outflowing component from among all the gas particles in the halo. Our definition is based on whether gas has enough kinetic energy to escape from the gravitational potential of the halo. Traveling from the inner part of the halo, the kinetic energy of the gas should satisfy the criteria below:
\begin{equation}
    \frac{1}{2}m v_{\rm esc}(r)^2 \geqslant \int_{r}^{R_{200}} \frac{G M(<r') m}{r'^2} dr' +  \frac{G M(<R_{200}) m}{R_{200}}
\end{equation}
Here $R_{200}$ is the virial radius of the halo, $M(<r)$ is mass of all types of particle enclosed within radius $r$ of the halo center. 
The first term accounts for the gravitation potential within the halo (as a gas particle move from $r$ out to $R_{200}$) while the  the second term the standard  component to escape from the "surface" of the halo.

The resulting escape velocity considers the fact that the gas from the inner region, in order to escape, needs to climb out the gravitational potential contributed by the rest of the mass within the halo, thus it is harder for it to escape than for the gas in outer region. This $v_{\rm esc}$ profile is hence different than simpler estimate of $v_{\rm esc}=\sqrt{2GM_{\rm halo}/R_{200}}$ as shown in Figure \ref{fig:figure2}. The black dashed line represents $v(r)=\sqrt{2}v_{\rm cir}=\sqrt{2GM(<r)/r}$, which is clearly below  our defined $v_{\rm{esc}}(r)$ in the inner region of the halo.  Making this
approximation (as was done by e.g., \cite{Barai2018}) may yield a higher estimate of the outflow gas (from the inner regions). 

\begin{figure*}

\begin{subfigure}[b]{\textwidth}
\includegraphics[width=\columnwidth]{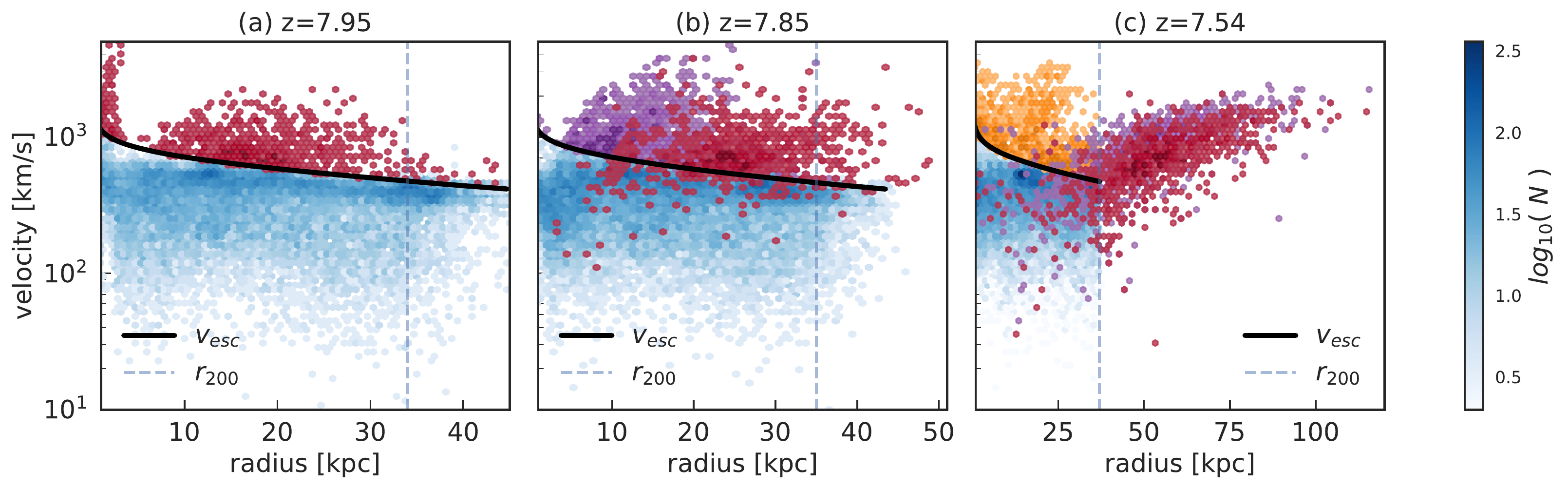}
\label{fig:b1}
\end{subfigure}

\begin{subfigure}[b]{\textwidth}
\includegraphics[width=\columnwidth]{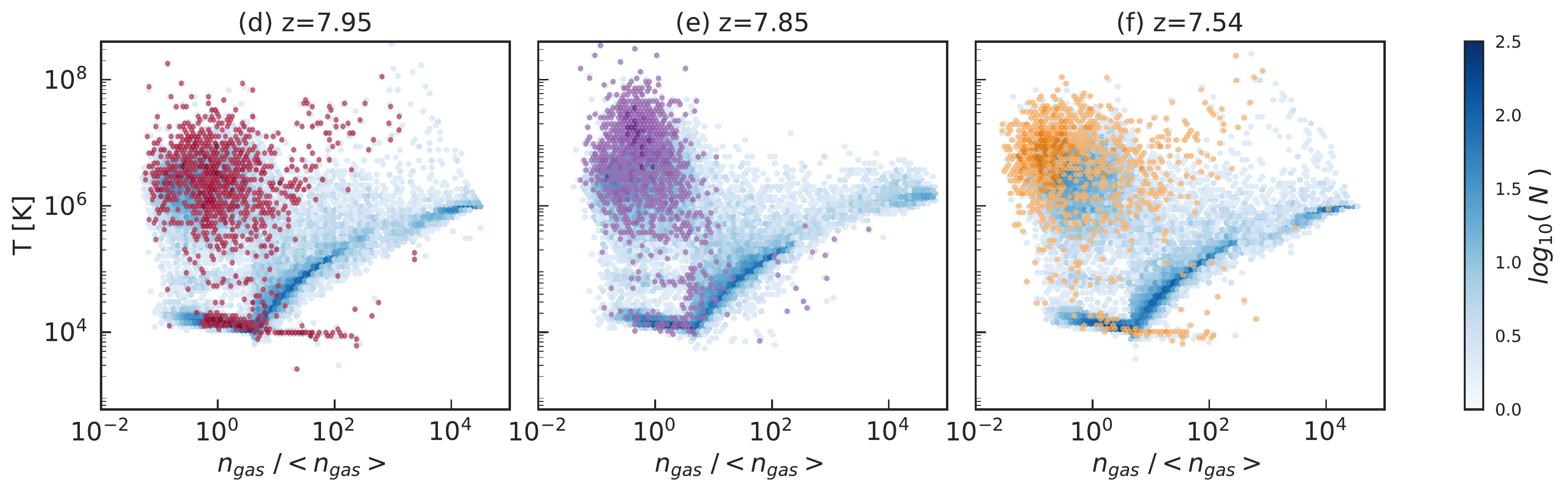}
\label{fig:b2}
\end{subfigure}

\caption{{\it Top row}: 2D histogram of gas velocity distribution versus the distance from the center of halo. The black line is the calculated escape velocity profile. The red points show the outflow particles identified at z=7.95 and their location in the corresponding $z=7.85$ and $z=7.54$ diagrams. The purple points show the new component of the the outflow at $z=7.85$, also shown at $z=7.54$. Finally, the orange points show the new outflow component at $z=7.54$.  
{\it bottom row}: the corresponding phase space diagram for the total gas and outflow gas. Note that the purple and orange dots now represent all the outflow gas, not only the new component in the outflow.}

\label{fig:figure4}
\end{figure*}

\begin{figure*}

    \begin{subfigure}[b]{0.3\textwidth}
        \includegraphics[width=\columnwidth]{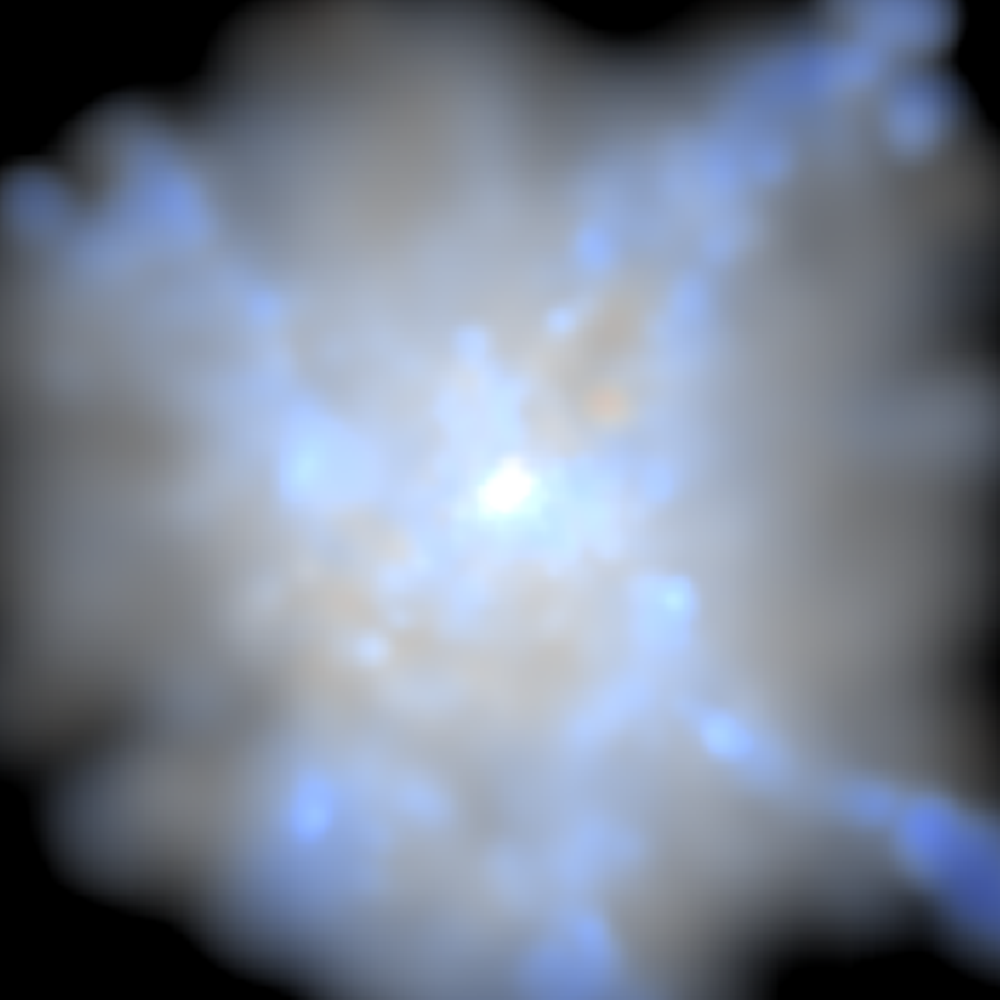}
        \caption{z=7.95}
        \label{fig:b1}
    \end{subfigure}
    ~  
    \begin{subfigure}[b]{0.3\textwidth}
        \includegraphics[width=\columnwidth]{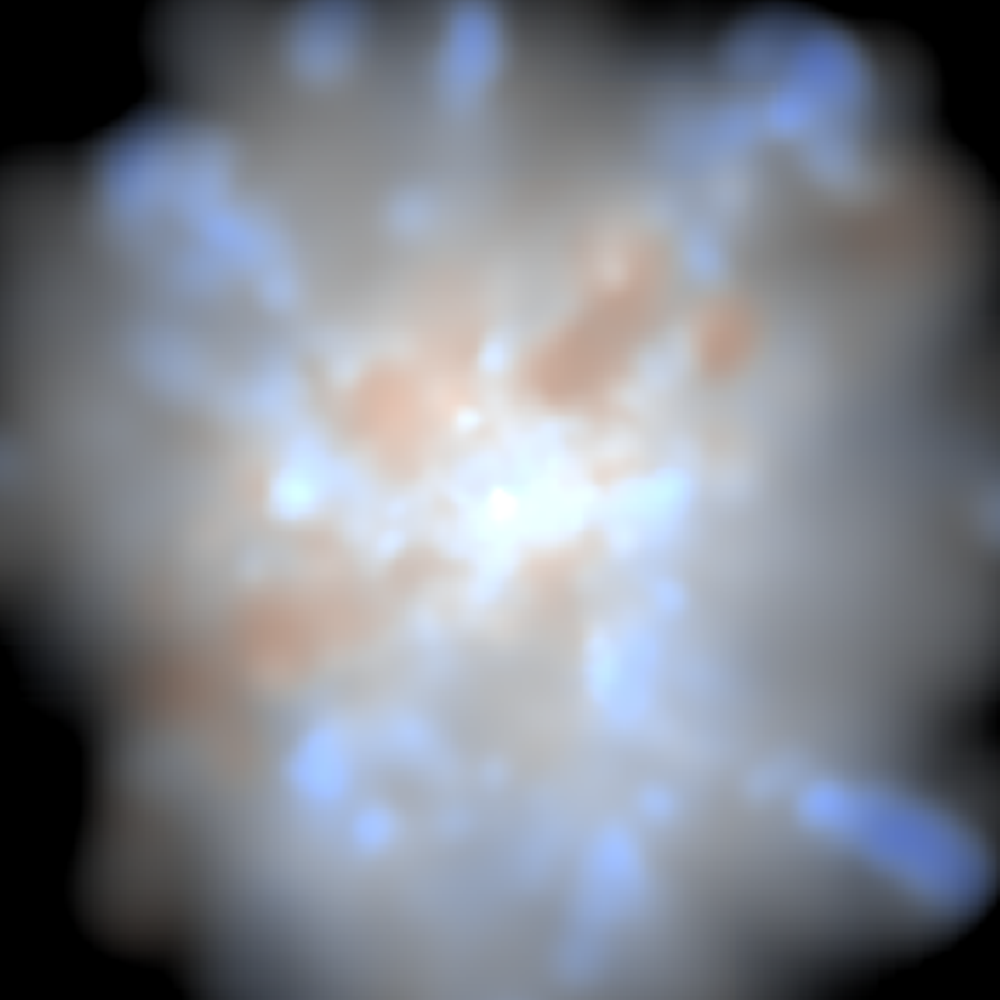}
        \caption{z=7.85}
        \label{fig:b2}
    \end{subfigure}
    ~
    \begin{subfigure}[b]{0.3\textwidth}
        \includegraphics[width=\columnwidth]{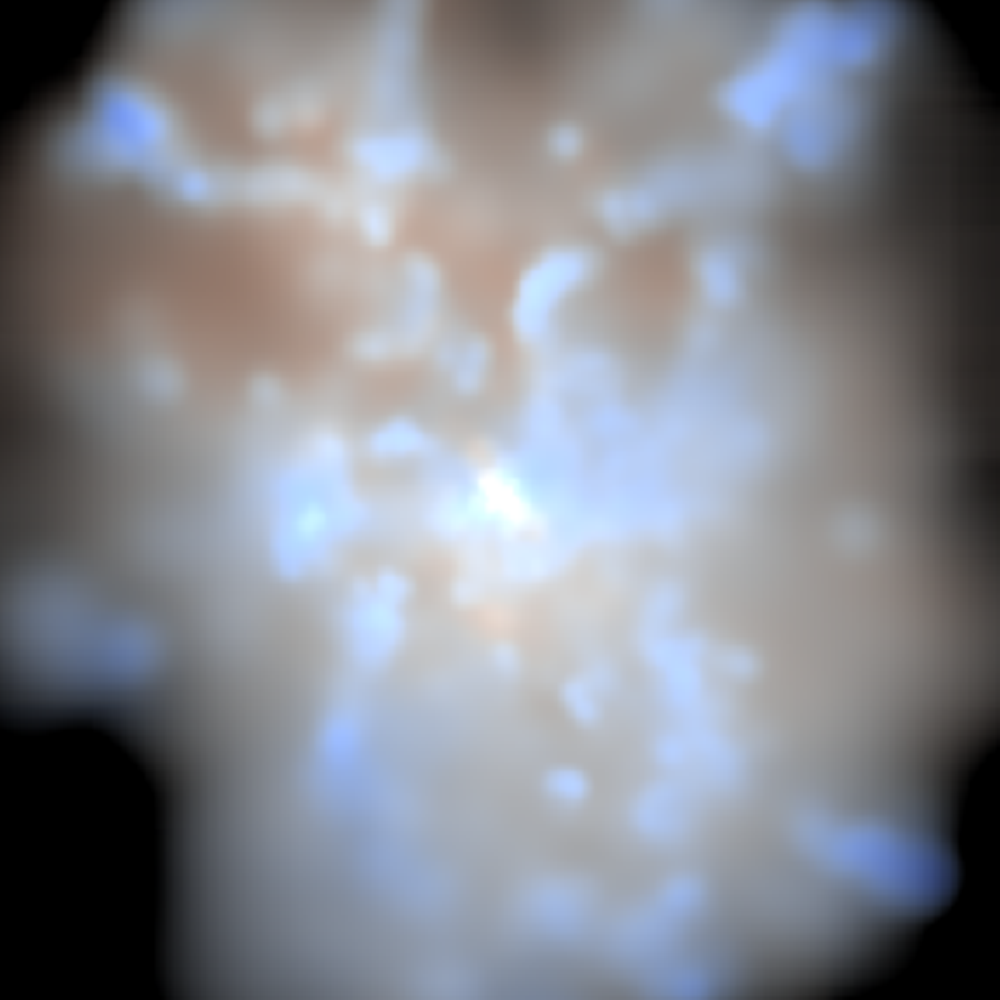}
        \caption{z=7.54}
    \end{subfigure}
    
    \begin{subfigure}[b]{0.3\textwidth}
        \includegraphics[width=\columnwidth]{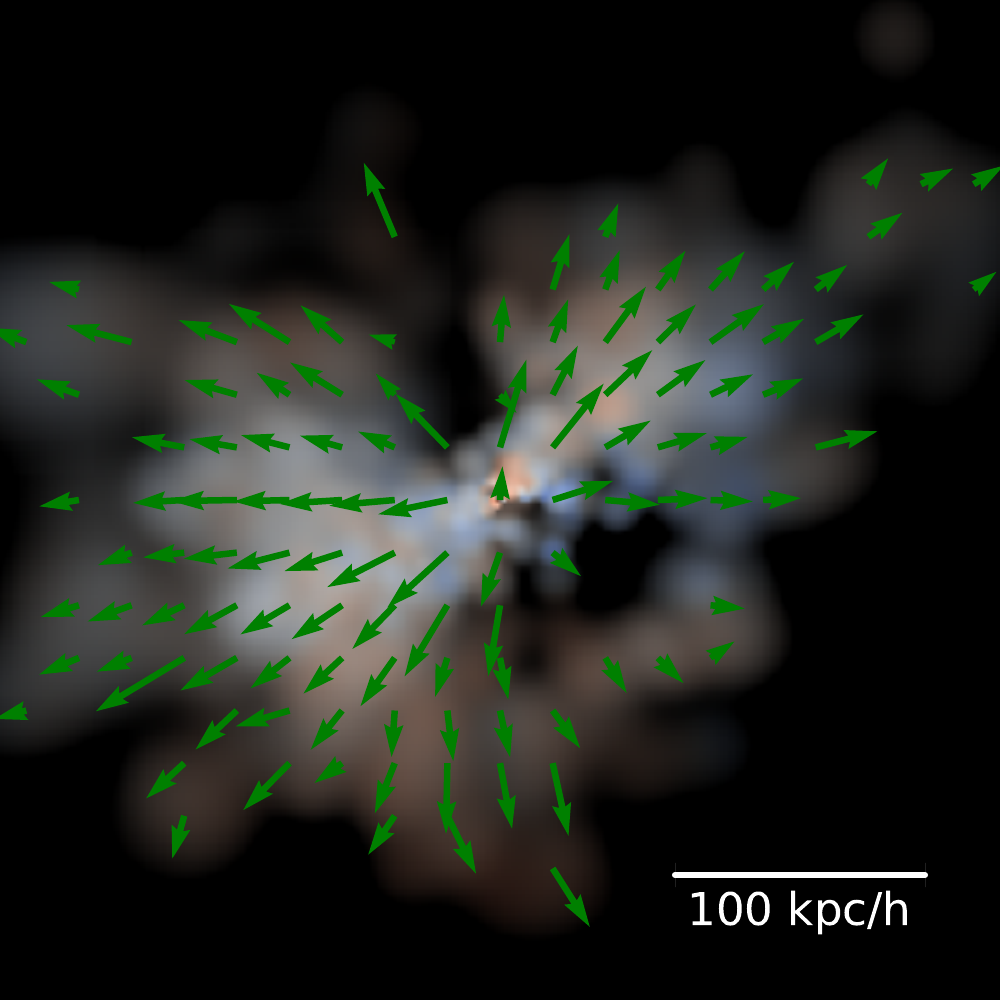}
        \caption{outflow gas, z=7.95}
    \end{subfigure}
    ~
    \begin{subfigure}[b]{0.3\textwidth}
        \includegraphics[width=\columnwidth]{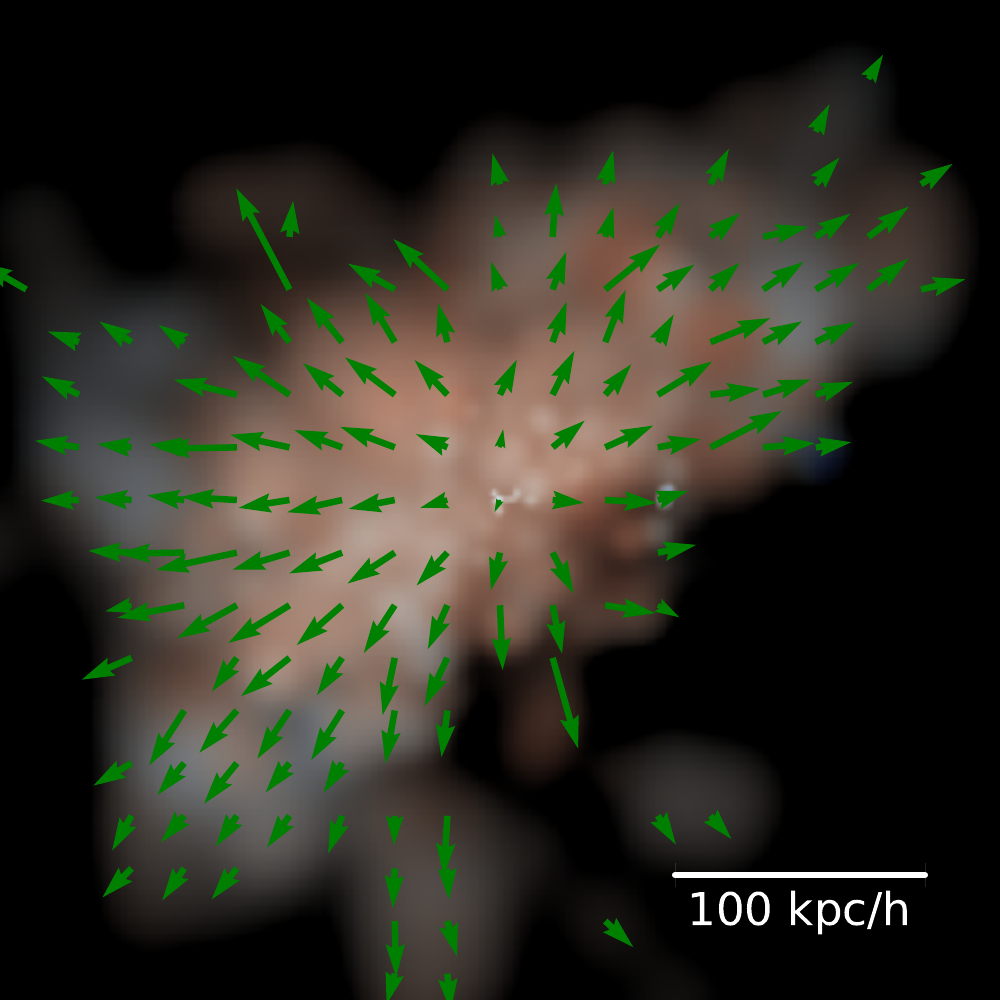}
        \caption{outflow gas, z=7.85}
    \end{subfigure}
    ~
    \begin{subfigure}[b]{0.3\textwidth}
        \includegraphics[width=\columnwidth]{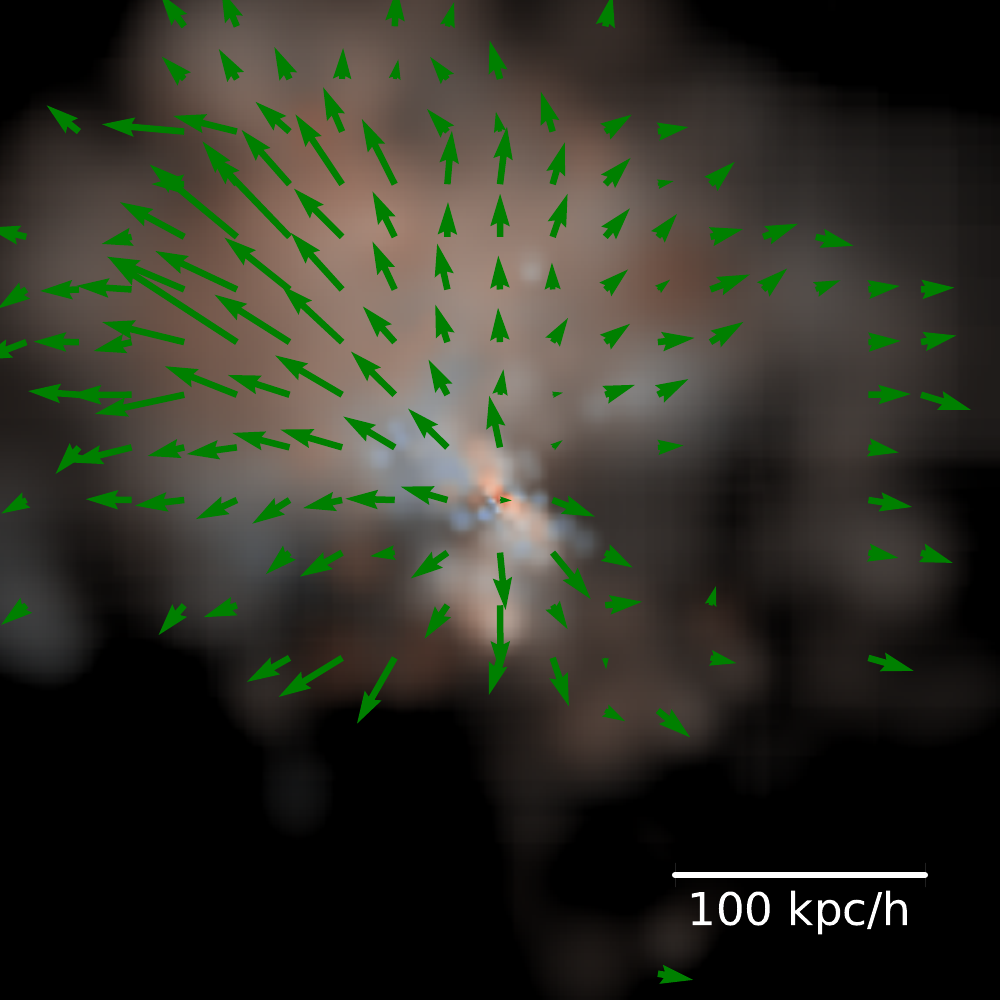}
        \caption{outflow gas, z=7.54}
    \end{subfigure}
        
    \caption{Images of the halo gas and outflow components in snapshots at different redshift, representing a $400h^{-1}$ kpc side (co-moving coordinate) box centered on the black hole. All panels show projection from the same direction, and are centered on the halo potential minimum (where the BH resides). The gas colour scale covers the  same range of density and mass averaged temperature in each panel.}
    \label{fig:figure5}
\end{figure*}

\begin{figure*}

    \begin{subfigure}[b]{0.4\textwidth}
 \includegraphics[width=\linewidth]{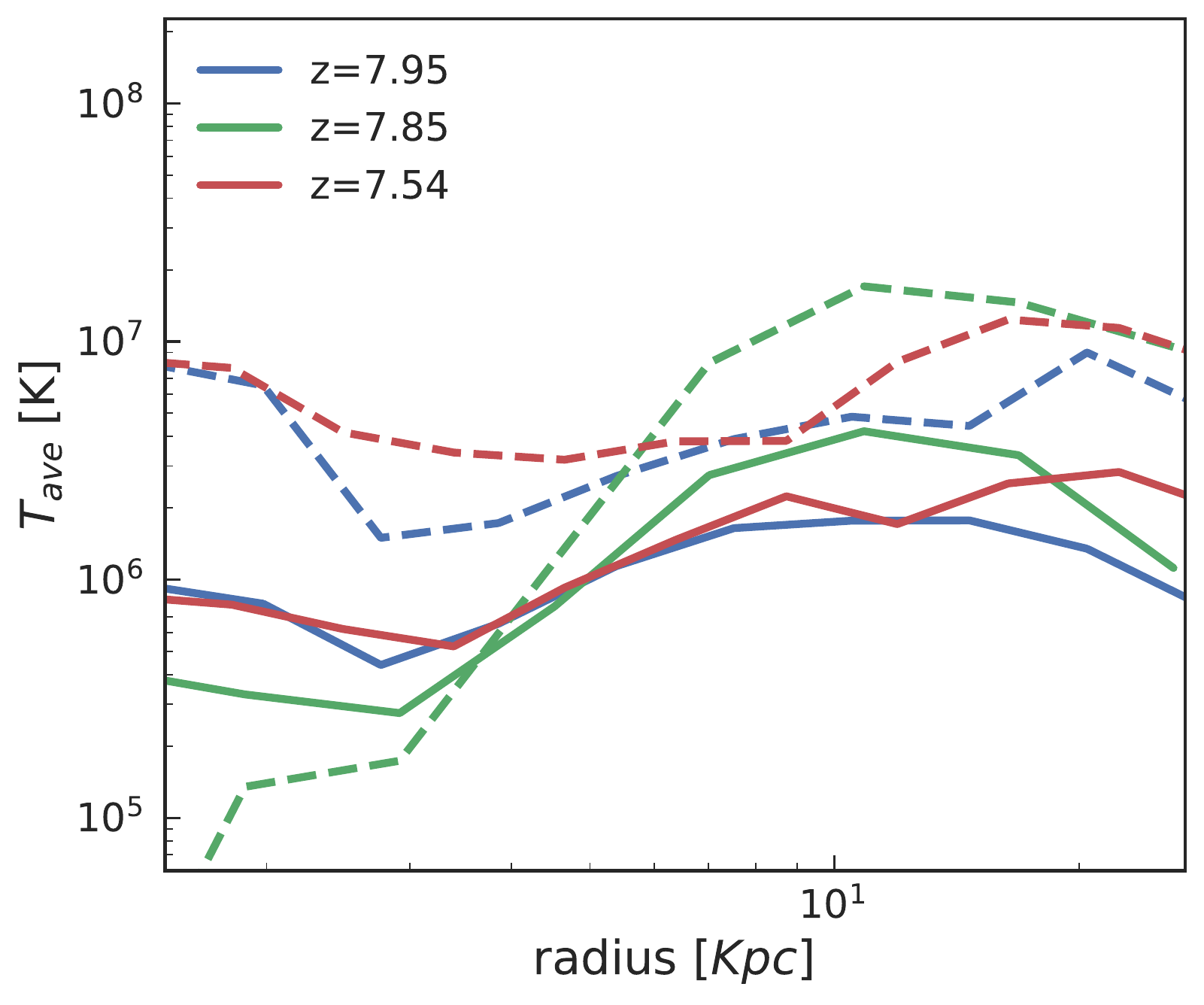}
        \label{fig:c2}
    \end{subfigure}
    ~
    \begin{subfigure}[b]{0.4\textwidth}
 \includegraphics[width=\linewidth]{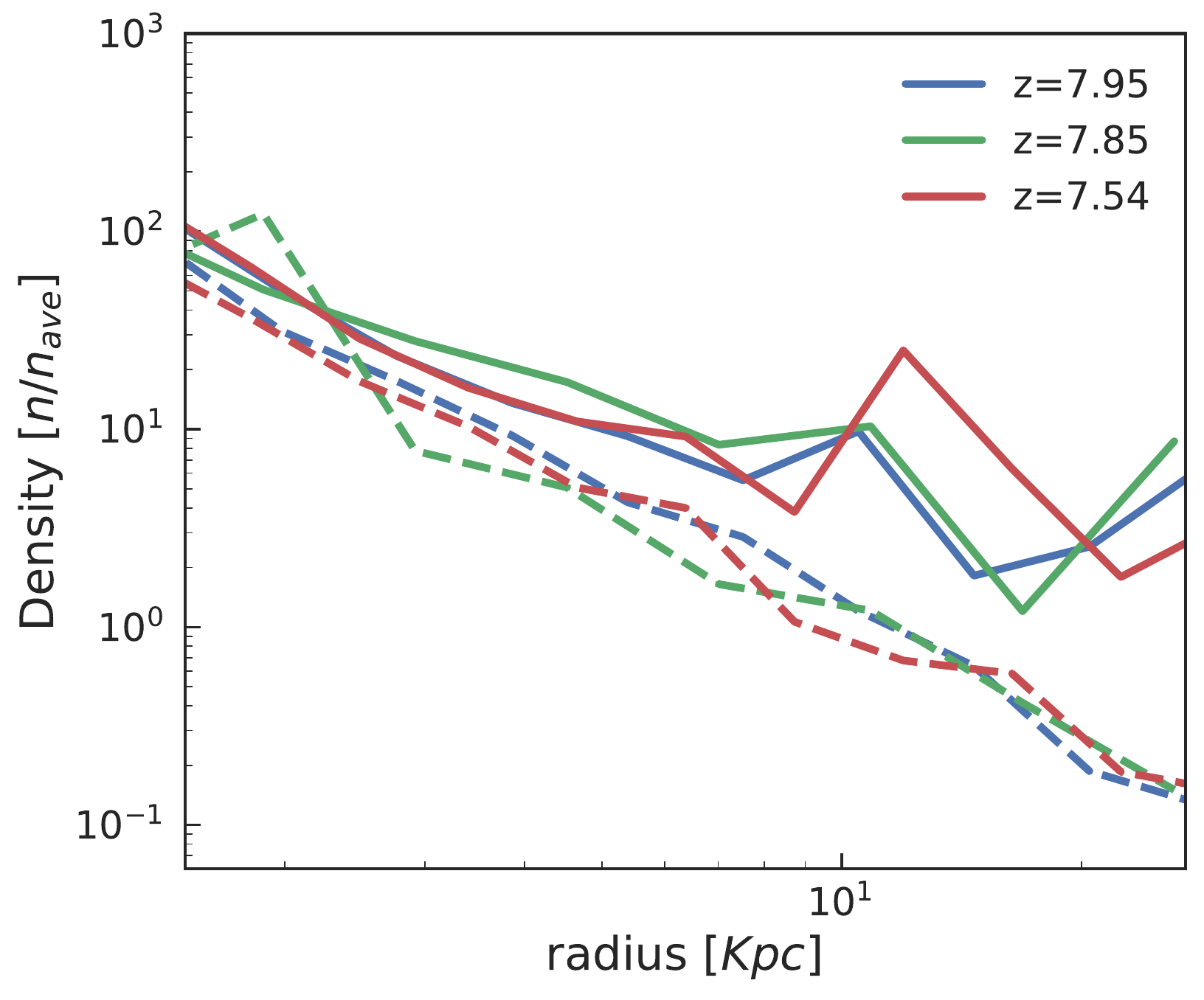}
        \label{fig:c3}
    \end{subfigure}
    \caption{The average temperature (left panel) and density (right panel) profiles of the gas (solid lines) and the outflow gas (dashed lines) within the halo. The dashed line represents the outflowing gas.}
    \label{fig:figure6}
\end{figure*}

\begin{table}
\caption{\label{molecular component} Total and outflow gas mass and molecular component}
\begin{tabular}{|c|c|c|c|}
\hline
 Redshift & z=7.95 & z=7.85 & z=7.54\\
\hline
$M_{\rm{H}_{2}}$ in halo $(\times 10^{10} M_{\odot})$ & 1.09 & 1.02 & 0.98 \\
$M_{\rm{H}_{2}}$ in outflow $(\times 10^{10} M_{\odot})$ & 0.022 & 0.013 & 0.026 \\
$M_{\rm gas}$ in halo $(\times 10^{10} M_{\odot})$ & 7.2 & 6.9 & 6.4 \\
$M_{\rm gas}$ in outflow $(\times 10^{10} M_{\odot})$ & 0.23 & 0.39 & 0.36 \\
\hline
\end{tabular}
\end{table}

\section{Results}

\subsection{Radial Velocities}
Figure \ref{fig:figure2} shows the calculated escape velocity profile  (solid line) along with the velocities the gas and dark matter particles as a function of radius (left and right panel respectively). Here we include only  particles which are in the FOF halo. As expected, 
dark matter particles (in the left panel), populate the part of the plot below the escape velocity
profile perfectly. If we compare this with the distribution of gas particles (in the right panel) we can see that there are many with velocities which exceed the escape velocity. Their high speeds are likely due to the effects
of AGN feedback, which leads to gas particles that can escape from the gravitational potential of the halo. \citep{Costa,Biernacki,Barai2018} 

Figure \ref{fig:figure3} shows the radial velocity histogram of gas particles in the halo. 
Each curve shows the mass fraction of gas per velocity bin, and the histogram is normalized to the total gas mass within the plotted region.
The gas with velocity magnitude $v > v_{\rm esc}$ (where $v$ includes the tangential component of velocity) is shown with a green solid line in Figure \ref{fig:figure3}. 
The red dashed line represents the molecular component in the gas with $v > v_{\rm esc}$. 
We define the outflow gas with radial velocity  ($v_{r} > 0$), corresponding to the region under the green solid line and to the right of $v_r=0$. 
We see that this outflow gas in the halo can have large outward radial velocity of over 3000 km/s.

The mass of gas participating in the outflow constitutes about 6\% of the total mass of gas in the halo, which is about $3.6\times10^{9}$${\rm M}_\odot$ at z=7.54 (see also Figure \ref{fig:figure01} and Table 1).
In particular, up 7\% of the gas in the outflow is in the molecular phase. This molecular gas typically has higher velocities (the peak of red histogram is at 1500 km/s) than the rest of the outflowing gas.  This is consistent with the observations of $z>6$ quasar outflows by ~\citep{Maiolino,Cicone2015}, where the cold gas traced by [CII] is moving with $v \sim 1400$ km/s.
We return to the molecular outflow component in \S 3.3.

\subsection{Outflow properties and their evolution}

In Figure \ref{fig:figure4}
we examine the formation, evolution and properties
of the outflows over the redshift range from $z\sim8$ to $z\sim7.5$.
In the top row we show the fate of the escaping gas particles and whether the outflow extends outside the halo. To do this we track the outflow gas particles through 3 snapshots of the simulation at $z=7.95, 7.85$ and $7.54$.  

Figure \ref{fig:figure4}(a) shows the gas particles in the halo at $z=7.95$, with the escaping, outflow component shown in red. At z=7.85, we find that about 55\% of the outflow gas from z=7.95 has escaped from the halo. The outflow particles retained in the halo are again shown in red while 
the new outflow is shown in purple. 70\% of the total outflow gas at z=7.85 is due to newly escaping gas particles. At $z=7.54$
we extend the radial distance probed and 
in (c), we show that indeed  most of the outflow gas has escaped from the halo: Only 7\% of outflow gas at z=7.95 and 10\% of outflow gas at z=7.85 still remains in the halo at z=7.54. Again, newly formed outflow (shown in orange) dominates the net outflow within the halo.
The net outflowing gas is shown
in the images in Figure \ref{fig:figure5}.

The bottom row panels of Figure \ref{fig:figure4} show
the phase space diagrams for the gas at the corresponding redshifts. We identify the outflowing gas using the same color coding as in the panels above. In each panel we show the temperature and density of the total outflowing gas within the halo at the specified redshift.
It is evident that the majority of the outflowing gas occupies the low density and high temperature part of the diagram.
However there is also a small fraction of colder and denser phase in the outflowing gas
(we discuss this further in \S 3.3).

In Figure \ref{fig:figure6} we show
the averaged temperature and density profiles of the halo gas and of the outflow at those redshifts. As seen in phase space, the temperature profile of outflow is higher on average than the rest of the gas.
The density profile of the outflow gas is steeper than the total gas in the halo, partly corresponding to the fact that the dense component in the outflow mainly resides in the inner region of the halo. 
Note also, that at $z=7.85$ the outflow gas in the inner parts of the halo is quite suppressed yet predominantly colder, (while the dominant component, warmer component is still in the outer region of the halos, see also \S 3.4).

\subsubsection{Morphology of the Outflows}
Figure \ref{fig:figure5} shows the morphology of halo gas and the outflow at $z=7.95,7.85$ and $7.54$. 
The top panels plot the gas particles in the FOF halo. 
The bottom panels represent the outflow component of the gas, showing both the density and the projected velocity field. 
The pictures are of a $400h^{-1}$ kpc side (co-moving coordinate) box centered on the black hole. 
All figures show the projection from the same direction.

In all panels of Figure \ref{fig:figure5}, the gas is coloured using the same scale which shows the density and mass averaged temperature. The yellow tint represents the warmer phase. We can see from the figure that the outflow gas is mainly composed of the hot phase together with a still small fraction of the colder and denser component. It is especially obvious in the plots of $z=7.85$ and $z=7.54$ that the outflow is closely tracing the morphology of the hot gas in the halo,

In the BlueTides simulation, the feedback energy from AGN is deposited in a sphere of surrounding gas.
However, we can see from the bottom panels of Figure \ref{fig:figure5} that the shape of the outflow can be highly anisotropic. 
The outflow has a bipolar shape at $z=7.95$, and can be more irregular as seen in the later redshifts, with the shape potentially varying a lot. One should remember that the outflow gas in each snapshot is physically distinct from the gas in the other snapshots (as shown in Figure \ref{fig:figure4} and discussed in \S 3.2), most of the outflow gas in the previous snapshot will have escaped from the halo by the next plotted redshift. 

\subsection{Molecular gas component in the outflow}

The gas cooling process in MP-Gadget is modeled based on radiative cooling and metal cooling. We include the molecular phase in the cold gas,  which is mainly hydrogen. The formation of hydrogen and its effect on star formation uses a semi-numeric model based on \cite{Krumholtz}, as described in \S 2.1.

The feedback processes in BlueTides include both SN wind and AGN feedback. As studied in \cite{Biernacki}, SN feedback creates a galactic fountain with dense gas clumps, enriching the metal component and boosts the cooling process. Meanwhile AGN feedback launches a low-density, hot outflow that sweeps through the extent of halo, pushing the dense clumps to large distances. 

Figure \ref{fig:figure2} presents the cumulative mass profile of stars and gas, and also the molecular component of the gas. At z=7.54, the mass of the molecular hydrogen in our halo is $9.8\times 10^9 \rm{M}_{\odot}$, consistent with ALMA observation of the host of the quasar at z=7.54~\citep{Venemans2017}, which give an upper limit for the molecular gas mass of $M_{\rm {H}_{2}} < 1.2\times 10^{10} \rm {M}_{\odot}$. 
Table~1 also provides the total gas and molecular gas mass within the halo at the three redshifts. The molecular gas fraction decreases slightly with decreasing redshift. 
Note that the molecular phase constitutes about 75\% of the mass fraction in the inner region of the halo within a radius of 3.3 kpc from the quasar, (corresponding to 20 $h^{-1}$ kpc in co-moving coordinate).
This is the part of the halo where most of the stars reside. The cumulative mass profiles of both stars and molecular component are nearly flat in the outer region of the halo, corresponding to the fact that star-forming gas mainly resides in the inner region of the halo.

We see from Figure \ref{fig:figure2} that most of the molecular component in the outflow gas is contributed by the outflow from the inner part of the halo. At z=7.54, the total mass of the outflow gas is $3.6\times 10^9 \rm{M}_{\odot}$, of which $2.6\times 10^8 \rm{M}_{\odot}$ is molecular hydrogen.

\subsection{Outflow rate}
\begin{figure}
\includegraphics[width=1.1\columnwidth]{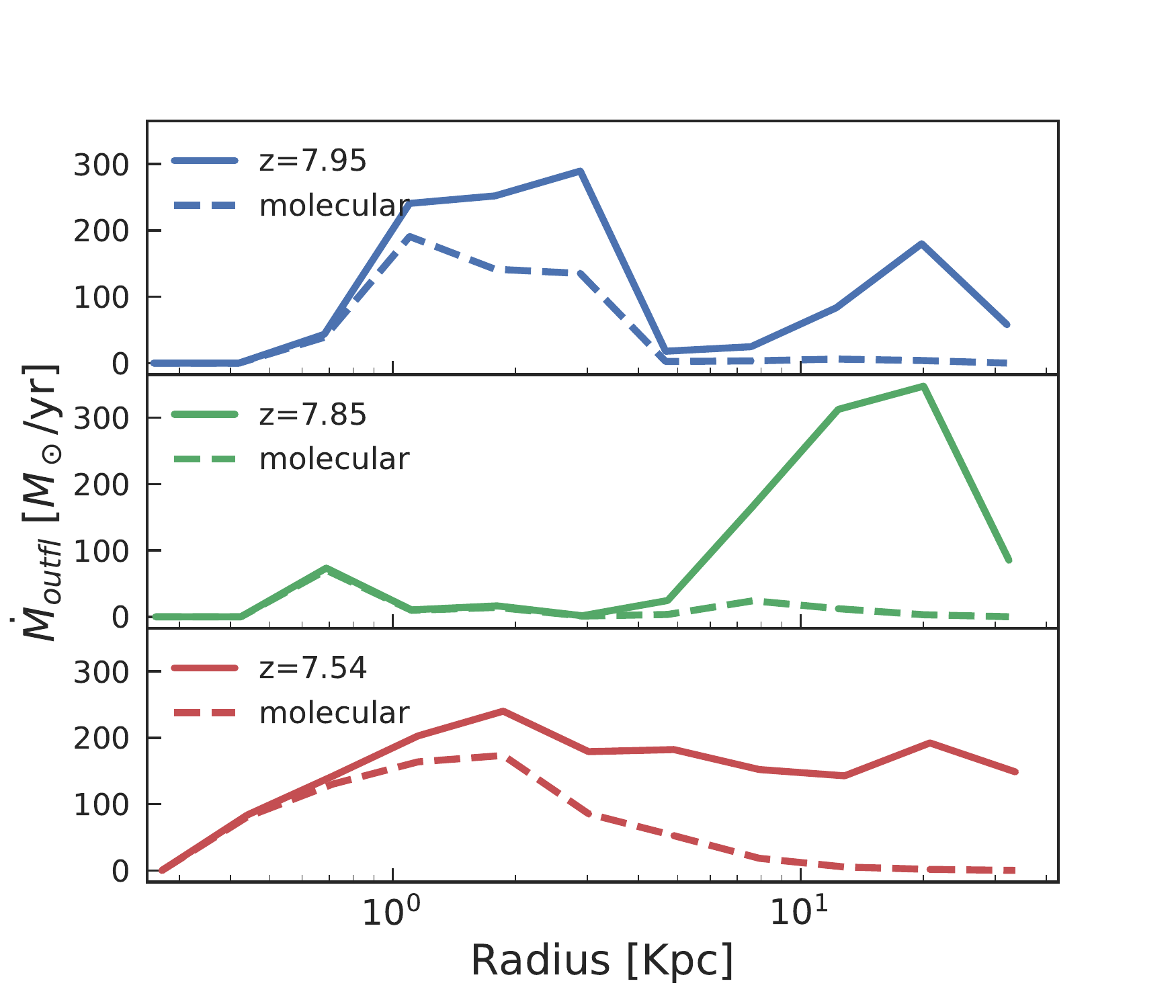}
    \caption{The mass outflow rate profile at different redshifts. The solid line shows all the outflow gas and the dashed line represents the molecular component in the outflow gas.}
    \label{fig:figure7}
\end{figure}

Figure \ref{fig:figure7} shows the outflow rate profile at different redshifts, using the outflow gas particles in our criteria. We calculate the outflow rate shell by shell to give the profile, using $\dot{M}_\mathrm{{outfl}(r)}=v_r\Omega r^2 \rho_{\mathrm{outfl}}$, with $v_r$ the radial velocity, and the volume averaged $\rho_{\mathrm{outfl}}=3M_{\mathrm{outfl}}/[\Omega(R^3_\mathrm{{out}}-R^3_\mathrm{{inner}})]$ being the average density of the shell. We note that our criteria for outflow are much stricter than the use of $v_r >0$ or $v_{\rm esc}=\sqrt{2GM_{\rm halo}/R_{200}}$, see Figure \ref{fig:figure2}. This makes our outflow rate quite conservative. 
Also, we note that the outflow rate becomes small when approaching  the edge of halo($\sim 36$ kpc) because we only count the gas particles "bounded" by the halo defined using the Friend of Friend algorithm based on dark matter particles. This will exclude some high velocity gas in the outer regions of the halo.

As shown in Figure \ref{fig:figure7}, the molecular component in the outflow extends into the region about 6 kpc from the halo center. This is consistent with the result of \cite{Biernacki} (see their Fig.9 middle column at 750Myr ($z\sim7$), noting that their plot is in co-moving coordinates while ours is in physical coordinates.) Their peak outflow rate for the dense gas component at $r=1.3$ kpc is about $50 \rm {M}_\odot/yr$, smaller than ours in the same region. However, our mass outflow rate varies over time. At z=7.85, the outflow is quenched in the inner region, probably corresponding to a quiescent state of the quasar.

\section{Conclusion}
We have studied the properties of the outflow gas driven by the most massive quasar in the BlueTides Simulation
in order to predict the outflow properties of the record holder $z=7.54$ recently discovered quasar. 

The epoch we have focused on is earlier than the $z\sim6$ studied by previous simulated work in this area ~\citep[e.g.][]{Costa, Curtis, Biernacki, Barai2018}.
Using our self defined escape velocity profile to select the outflow gas that can escape the gravitational potential of the halo, we find the total mass of the outflow gas to be $3.6\times 10^{9} \rm{M}_\odot$, constituting 6\% of the total gas in the halo. 
Furthermore, the outflow gas contains a cold, dense molecular component with mass about $2.6\times 10^{8}\rm {M}_\odot$. The molecular phase of the gas mainly resides in the inner region of the halo,  in the star forming region within a few kpc of the central black hole. 

The outward radial velocity of the gas particles in the halo can reach upwards of $3000$ km/s, within which the molecular component typically has velocities higher than the rest of the outflow gas, with mass averaged $v_r \sim 1300$ km/s. These values are consistent with the observations of \cite{Cicone2015}, who detect a cold dense molecular gas outflow with $v_r \sim 1400$ km/s in a quasar at $z\sim 6$ implying that the properties
of the outflows around quasars at $z > 7$ may be rather similar to the population of $z\sim 6$ quasars.

The outflow gas can escape from the halo in a short period of time. More than 90\% of the outflow at z=7.85 has escaped from the host halo at the later redshift of z=7.54 (40 Myr later). 
Based on our criteria for outflow, the average mass outflow rate is typically $200-300 \rm {M}_\odot/\mathrm{yr}$.

The outflow gas is mainly composed of a hot and sparse phase. However we find that there is indeed a small fraction ($\sim 10$\%) of outflow gas in a cold (molecular) and dense phase.
On average, the outflow has higher temperature($T \sim 10^7 $K) and lower density compared to rest of the gas in the host halo, and has a steeper density profile. The dense molecular fraction resides in the inner halo. In a companion paper \cite{Tenneti2018} studying the host galaxy of this simulated quasar, we show that is likely to contain large amounts of dust. Already in \cite{DiMatteo2017} we saw that this galaxy had a major episode of star formation at $z\sim 10$
(reaching $\sim 300  {\mathrm M}_\odot/\mathrm{yr}$, see the green line in Figure of that paper). Both of these features are consistent with the observations of \cite{Venemans2017}, and we predict that the outflows we have seen in the simulated galaxy are likely to be present in the observed quasar. In addition, the presence of such signicant quasar driven outflows may help explain the significant metal enrichment that the quasar at $z> 6$ host have been observed to have. The quasar outflows can easily spread the metal enriched 
gas over and beyond the scale of the halo.

\section*{Acknowledgments}
We acknowledge funding from NSF
ACI-1614853, NSF AST-1517593, NSF AST-1616168, NASA ATP NNX17AK56G and NASA ATP 17-0123 and the BlueWaters PAID program. The \texttt{BLUETIDES} simulation was run on the BlueWaters facility at the National Center for Supercomputing Applications

\bibliographystyle{mnras}
\bibliography{bib.bib}




\bsp	
\label{lastpage}

\end{document}